\pdfoutput=1
\documentclass[epj]{svjour}
\usepackage[english]{babel}
\usepackage[activate=normal]{pdfcprot}
\usepackage[hidelinks]{hyperref}
\usepackage{amsmath,amssymb,amsfonts,latexsym,bbm}
\usepackage{graphics,color,url}

\newcommand{\bfr}{{\mathbf{r}}}

 \newcommand{\calC}{{\mathcal{C}}}

 \newcommand{\calF}{{\mathcal{F}}}

 \newcommand{\calP}{{\mathcal{P}}}

 \newcommand{\calS}{{\mathcal{S}}}
 \newcommand{\calT}{{\mathcal{T}}}

\newcommand{\Reals}{\mathbbm{R}}

\newcommand\scriptbox[2]{{\makebox[#1]{\hss$\scriptstyle #2$\hss}}}
\newcommand\scriptboxl[2]{{\makebox[#1]{\rlap{$\scriptstyle #2$}\hfill\hbox{}}}}

\def\cA{\bfr^A}
\def\cB{\bfr^B}
\def\takeout#1#2{#1\backslash #2}

\newenvironment{algo}%
{\smallskip\begin{list}{}{%
  \sffamily
  \rightmargin=0pt%
  \itemsep=0pt \itemindent=0pt \listparindent=\parindent%
  \partopsep=0pt \parskip=0pt \parsep=0pt%
  \labelsep=0.5em \topsep=0.2em
  \labelwidth=0.5em%
  \leftmargin=\labelwidth \addtolength{\leftmargin}{\labelsep}%
}}%
{\end{list}\smallskip}

\newcommand{\fv}{\textsc{fv}}
\newcommand{\av}{\textsc{av}}
\newcommand{\avato}{\textsc{avato}}
\newcommand{\avatoa}{\textsc{avato-a}}
\newcommand{\avatob}{\textsc{avato-b}}

\begin{document}
\title{Cavity averages for hard spheres in the presence of polydispersity and incomplete data}
\titlerunning{Cavity averages in the presence of polydispersity and incomplete data}
\author{Michael Schindler \and A.~C. Maggs}
\institute{UMR Gulliver 7083 CNRS, ESPCI ParisTech, PSL Research University, 10~rue Vauquelin, 75005~Paris, France}
\date{\today}
\abstract{
We develop a cavity-based method which allows to extract thermodynamic
properties from position information in hard-sphere/disk systems. So far, there
are \emph{available-volume} and \emph{free-volume} methods. We add a third one,
which we call \emph{available-volume-after-takeout}, and which is shown to be
mathematically equivalent to the others. In applications, where data sets are
finite, all three methods show limitations, and they do this in different
parameter ranges. We illustrate the principal equivalence and the limitations
on data from molecular dynamics -- In particular, we test robustness against
missing data. We have in mind experimental limitations where there is a small
polydispersity, say 4\% in the particle radii, but individual radii cannot be
determined. We observe that, depending on the used method, the errors in such a
situation are easily~100\% for the pressure and~$10\,kT$ for the chemical
potentials. Our work is meant as guideline to the experimentalist for choosing
the right one of the three methods, in order to keep the outcome of
experimental data analysis meaningful.
}
\maketitle

\section{Introduction}
Recent years have seen a growing number of experiments which use real-space
measurements to extract quantitative information on thermodynamics and on
dynamics of
colloids~\cite{AndLek02,BlaWil95,KegBla00,RoyPooWee13,YunCheGraLohStiYod14}.
Some of the investigated systems are
crystalline~\cite{ZahWilMarSenNie03,StiGooCheYunLiuYod14,LiuBessHerDemImhBla14,TafWilTanRoy13,GhoMarChiSchMagBon11,BauDulDijRotBec07},
some are disordered and more or less
dense~\cite{DreXuStiHouYodTor15,ZarRusSchTanBon14,ZarNieSchBon13,KliEbeWeyFucMarKei12,ChiSch12,GhoChiSchBon11}.
Extracting a full thermodynamic description out of the available data is still
a challenging task. Typical data consist of configuration snapshots, for
example from video recording, or from confocal microscopy, which provide the
positions of all particles at given times. Any thermodynamic description
depends on the choice of the disregarded degrees of freedom. In particular, one
typically disregards the degrees of freedom of a surrounding fluid and
describes it in a more or less effective way. The challenge in this task thus
consists in the determination of a set of thermodynamic variables, say pressure
and chemical potentials, which is consistent, that is both quantities are
calculated from the same data -- for example from the particle centers alone,
disregarding surrounding fluids.

A number of experiments aim at hard
colloids~\cite{RoyPooWee13,SchMarMegBry07,ZarNieSchBon13}, where the data
analysis is the same as for hard-sphere simulations. There, simulations can
serve for the development of good data treatment algorithms and for their
calibration. On top of the conceptual question how to extract a consistent set
of thermodynamic quantities, in real-world application, also the robustness of
these algorithms against noise and missing data is important. Usual sources of
noise in experiments are limited resolution of position and size of the
particles~\cite{PooWeeRoy12}, shape variations, and many more. In particular,
there seems to be an unavoidable amount of size variations (polydispersity) of
around 4\%~\cite{SchMarMegBry07,ZarNieSchBon13}. Are the used algorithms for
the data analysis robust against this variation?

\begin{figure}%
  \centering
  \includegraphics{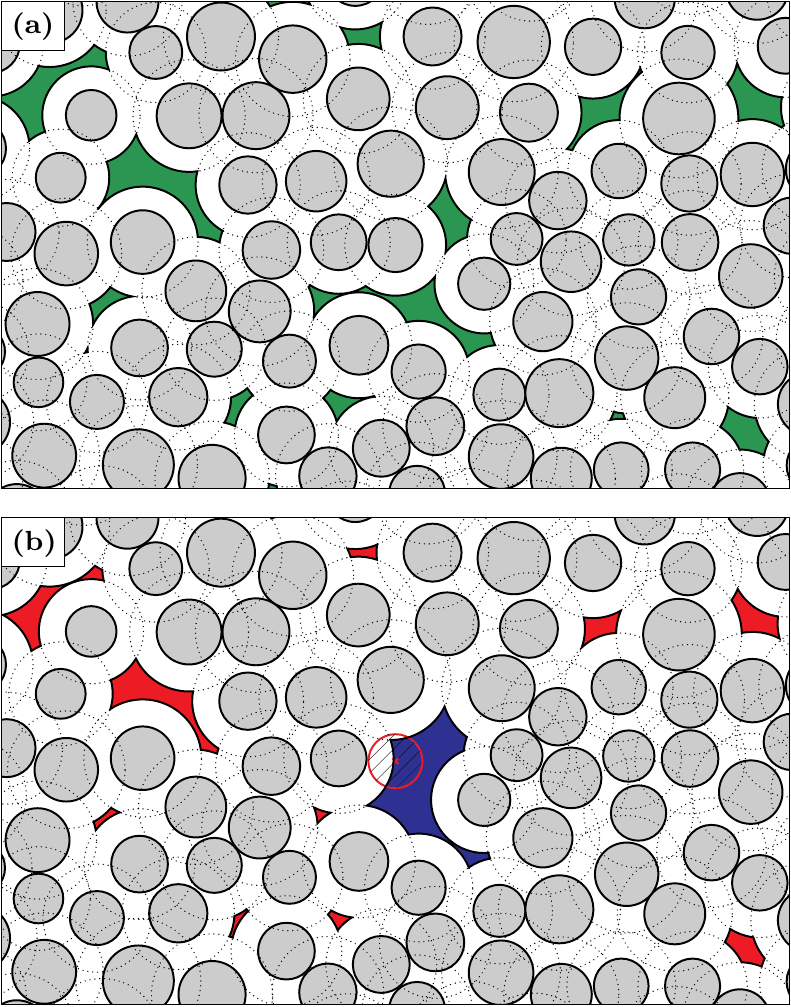}%
  \caption{Some cavities in a given polydisperse configuration of hard disks.
  Dotted circles indicate the excluded volume for the center of the particle in
  question. \textbf{(a)}~Available volume (in green) for an additional
  particle. \textbf{(b)}~Free volume (blue) of the indicated particle
  (hatched). The new \avato{} average uses the union of the free volume and the
  remaining cavities (red).}%
  \label{fig:av_fv}
\end{figure}%
For hard spheres the determination of pressure and chemical potentials reduces
to a geometrical problem. One family of algorithms to calculate them is based
on measuring the space where a particle can be inserted into a given
configuration. The general scheme is to take a set of configurations, calculate
a certain geometrical quantity from each, and average it over the
configurations. The different methods vary in the choice of the geometrical
quantity and of the configurations. Widom~\cite{Widom63} described the
\emph{available volume}, that is the volume~$V_0$ where yet another particle
can be inserted into a configuration of $N$~particles. Clearly, the available
volume may consist of several disconnected regions \emph{(cavities).}
Figure~\ref{fig:av_fv}a shows an example, where $V_0$~is the union of the green
volumes. The notion of chemical potentials is directly linked to the insertion
of another particle into a system. The link between $V_0$ and the pressure is
less obvious. It has been established by Speedy~\cite{Speedy80} who expressed
the pair-distribution function~$g(r)$ in terms of the ratio of
averages~$\langle S_0\rangle/\langle V_0\rangle$, where the average is done
over all possible configurations and $S_0$ is the area of the surface of~$V_0$. This work
made it principally possible to calculate the pressure from cavity averages. On
a different line of reasoning, Hoover \emph{et al.}~\cite{HooAshGro72}
expressed the pressure on the basis of the region which a particle can explore
when all others remain fixed. This cavity, which is called \emph{free volume},
is another geometrical quantity which can be extracted from a given
configuration.\footnote{There is another branch of cavity-based method which
goes back to Kirkwood~\cite{Kirkwood50} and which has been specialized to hard
spheres by Wood~\cite{Wood52}. This method also uses the term \emph{free
volume}, but it should not be confused with the free-volume method presented
here. The essential difference is that in the Wood/Kirkwood method, the
cavities are extracted from one averaged snapshot, where the averaging may have destroyed the
non-overlapping constraint of hard spheres. Here, cavities are taken from many
snapshots which are all compatible with the constraint, and the average is done
subsequently.} An example is given in fig.~\ref{fig:av_fv}b. Notice that if a
different particle is chosen, the resulting free volume is different, and the
free volumes of two particles may overlap. Clearly, the free volume has
advantages over the available volume in dense systems, where adding yet another
particle might be impossible. The free volume, however, never vanishes, because
we first take out the particle in question, then determine the cavity where we
can put it back in. Hoover \emph{et al.} used a dynamical argument for their
formula for the pressure. Speedy~\cite{Speedy81} found the same formula by
showing that the above ratio of averages equals an average of a ratio, $\langle
S_f/V_f\rangle$, where $V_f$~is the free volume, $S_f$~its boundary. Some years
later, Speedy~\cite{Speedy88,SpeRei91a} found a way to re-derive the pressure
result without recourse to the pair distribution function, by only counting
configurations and applying elementary thermodynamics. This approach is more
precise on how the averages are calculated. Interestingly, in order to get the
pressure of $N$~particles, one needs averages over ensembles of
\hbox{$(N{-}1)$} spheres. This difference is the key for the introduction of
free-volume methods, where configurations of one particle less are regarded.
Taking this difference seriously allows to provide algorithms which work
seamlessly in all cases, being dilute, dense disordered, or crystalline.

If the particles have different sizes, the precise way how to average over
cavities changes. Instead of a single available volume, we now have to
calculate one for each radius of particle we want to insert. Finally, pressure
and chemical potentials are weighted averages over these available volumes. How
to calculate this average was shown by Corti~\& Bowles~\cite{CorBow99}. They
generalised Speedy's work that uses pair-distribution functions to the
polydisperse case. They did not use the cleaner configuration counting
approach.

In the present paper we enrich the family of cavity-based algorithms by one
member. To the methods based on \emph{available-volume}~(\av) and on
\emph{free-volume}~(\fv), we add a third one, which we call
\emph{available\discretionary{-}{}{-}volume\discretionary{-}{}{-}after\discretionary{-}{}{-}takeout}
(\avato). Its derivation below makes the \avato{} method appear as the natural
extension of the configuration counting idea, when passing from averages over
\hbox{$(N{-}1)$} particles to those over $N$~particles -- similarly to the
free-volume method, but conceptually simpler. It is mathematically equivalent
to the other two methods if averages over all configurations are available. As
does the \fv~method, also the \avato-average is possible in dense
configurations. We provide formulae and algorithms for calculating both the
pressure and the chemical potentials in all three methods, in the presence of
polydispersity. For the two established methods, \av~and \fv, most of the
equations have already been published and
implemented~\cite{SasCorDebSti97,SasTruDebTorSti98,MaiLakSas13,MaiSas14} -- but
not for all the combinations of chosen method, pressure, chemical potential and
polydispersity. For completeness we give all the equations here.

The second contribution of the present paper is to apply the derived algorithms
to numerical data in two dimensions. We reproduce their principal equivalence
and show how they start to differ on finite sets of data. We further test their
robustness if we throw away information on individual particle radii at small
values of polydispersity. The provided comparisons should guide
experimentalists working with colloids, such that they can avoid the indicated
problems and choose the best algorithm for analysing their data.

The outline of the paper is as follows. In sects.~\ref{sec:av} and~\ref{sec:fv}
we recall the established cavity methods and define some notation, restricting
ourselves to the monodisperse case. The \avato~average is derived
in~sect.~\ref{sec:avato}. The expressions for pressure and for chemical
potentials in the fully polydisperse case follow in sects.~\ref{sec:press}
and~\ref{sec:mu}. Numerical results of all three methods are presented for
pressure (sect.~\ref{sec:num:p}) and for chemical potentials
(sect.~\ref{sec:num:mu}). The effect of missing radius information is shown in
sect.~\ref{sec:num:missing}.

\section{Cavity methods}
We now present the established cavity methods and derive our new method in the
monodisperse case. We will also define the necessary ensembles and averages.
Consider a collection of $N$~spheres/disks of diameter~$\sigma$ in
$d$~dimensions, enclosed in a volume~$V$ at temperature~$T$. In order to
exclude surface effects, we take the volume to be periodic.
\subsection{Available volume}
\label{sec:av}
Speedy~\cite{Speedy88} arrived at an expression for the pressure
in the monodisperse case, using only counting arguments. His result is the
ratio of two averages,
\begin{equation}
  \label{mono:p:av}
  \frac{pV}{NkT} = 1 + \frac{\sigma}{2d}
    \frac{\bigl\langle S_0(k)\bigr\rangle_{k\in\Omega(N{-}1)}}
         {\bigl\langle V_0(k)\bigr\rangle_{k\in\Omega(N{-}1)}}.
\end{equation}
Here, $\Omega(N{-}1)$ is the set of all possible ways to place
$(N{-}1)$~spheres in the volume, and $\langle\cdot\rangle_{\Omega(N{-}1)}$~is
the average over this set. In order to render the set finite, we may think of
space being cut into small pixels of volume~$\omega$ and observe that the final
expression for the pressure does not depend on~$\omega$. We denote one such
configuration (or ``state'') by the centers $\bfr_i$ of the spheres,
$k\in\Omega(N{-}1): k=\{\bfr_1,\dots\bfr_{N{-}1}\}$. Despite every particle
having a unique number, they are not discerned within a state. The
\emph{available volume}~$V_0(k)$ is then the volume where the center of an
additional particle, $\bfr_N$~in the above case, can be placed. See
fig.~\ref{fig:av_fv}a for an example. The available volume may fall into
several disjoint connected regions, called \emph{cavities}. $S_0(k)$~is the
boundary of the available volume.

The formula~\eqref{mono:p:av} is not directly applicable in data analysis, for
two reasons: First, if we have data from an experiment or a simulation with
$N$~particles, we never have access to strictly all configurations. We rather
take a finite series of $K$~snapshots, $\calS(N)$, and average over those.
$\calS(N)$~represents the data which is really available from an experiment.
The number of snapshots, $K$, introduces a first level of approximation, which
becomes exact in the limit $K\to\infty$. We will therefore denote this
approximation as an equality. The second problem is that we have snapshots of
$N$~particles, not of $(N{-}1)$~particles. This problem gives rise to the
\emph{free-volume} and the \emph{available-volume-after-takeout} methods below.
Here, it simply introduces another level of approximation.

Applying eq.~\eqref{mono:p:av} to real-world data, we approximate the averages
by those which we have, namely over the snapshots $\calS(N)$, which contain all
particle centers (and radii) at given times,
\begin{equation}
  \label{mono:p:AV}
  \frac{pV}{NkT} - 1 \approx \frac{\sigma}{2d}
    \frac{\bigl\langle S_0(k)\bigr\rangle_{k\in\calS(N)}}
         {\bigl\langle V_0(k)\bigr\rangle_{k\in\calS(N)}}.
\end{equation}
Equation~\eqref{mono:p:AV} can be translated into the following algorithm:\\
\begin{minipage}{\columnwidth}
\begin{algo}
\item Loop over the snapshots~$k\in\calS(N)$:
  \begin{algo}
  \item Enlarge all sphere diameters by~$\sigma$.
  \item Find the space not covered by any sphere (cavities).
  \item $V_0(k) \leftarrow$ sum of all cavity volumes.
  \item $S_0(k) \leftarrow$ sum of all cavity boundaries.
  \item Accumulate $S_0(k)$ and $V_0(k)$ in independent averages.
  \end{algo}
\item End loop over~$k$.
\end{algo}
\end{minipage}
The enlargement step takes care of the excluded volume of both the present
particles and the additionally inserted one, see fig.~\ref{fig:av_fv}. The
cavities are for the \emph{center} of the inserted particle only. The algorithm
to find the volumes and boundaries which are not covered by any sphere is
discussed in sect.~\ref{sec:nummethods}.
\subsection*{Effect of finite $N$ and $K$}
\label{sec:finite}

Generally speaking, in the limit $N\to\infty$ and $K\to\infty$
equation~\eqref{mono:p:AV} and similar expressions in the following sections
become exact. In practice both parameters are finite, and this introduces
deviations in the approximations. The precise nature of the deviations is
subtle, they depend on the number density~$N/V$ and on the quantity that is
being averaged. We now try to capture some aspects of the deviations, without
being exhaustive.

Let us focus on the individual effect of finite~$N$ first, assuming $K$~to
cover all possible configurations. This is as if we replaced $\calS(N)$ by
$\Omega(N)$ in eq.~\eqref{mono:p:AV}. Still, this equation aims at calculating
the pressure of a different system than the original one, namely $p(N{+}1,V,T)$
instead of~$p(N,V,T)$. There is a (small) error in the density of the order
$1/N$ which translates into an error in the pressure. We can neglect it in the
limit of many particles.

Now, if $K$~is finite, everything depends on the concrete configurations
contained in~$\calS(N)$. Above all we need the cavities to be sufficiently
numerous to build reliable averages. If the particle density is sufficiently
small, this is the case because we can nearly always insert another particle.
The density deviation described above is thus the only systematic error and can
be controlled by choosing $N$~large enough. The finite number of snapshots
introduces additional random noise. Surely, $K\to\infty$ will make this noise
disappear, but this does not imply that the rate of convergence is sufficient
for practical applications. This point will remain open in the present paper;
we would only like to mention that there are examples in the
literature~\cite{CheStiSchAptSchMagLiuYod13} where the quantity of interest
converges so slowly that advanced extrapolation methods are required.

The influence of finite~$K$ and $N$ is more subtle for crystalline systems. It
may happen that not a single snapshot in the given set~$\calS(N)$ allows
insertion of another particle. In such a case we cannot calculate the
average~$\langle\cdot\rangle_{k\in\calS(N)}$. (What we said above in the exact
limit $K\to\infty$ remains valid, however, because we will find at least one
extensible configuration in $\Omega(N)$ if $N$ is large enough for the given
density.) In order to see cavities in a monocrystal of $N$~particles, we rely
on \emph{spontaneous} fluctuations to make sufficient space for particle
insertion. Their rate of appearance is a function of the density $N/V$ and of
the total number~$N$ and becomes exponentially small at high
densities~\cite{SpeRei91a,BowSpe94}. In practice, where $K$~is limited to a few
thousand, we are not astonished to see few or no cavities in the snapshots and to
see badly converged averages.

For computer simulations one way out of this problem is to take $\calS(N{-}1)$
from the beginning, that is to simulate a crystal with a
vacancy~\cite{SpeRei91b}. This guarantees that the number of cavities is around
one even in the densest system. In fact, this gives directly an approximation
to eq.~\eqref{mono:p:av}. We will not pursue this idea further in the present
paper, because it is rather limited to computer simulations, and we here focus
on what can be done with experimental data, taking simulations only as a test
ground. Notice however that the number of cavities, $N_c(k)$, is a nice example
for a function that does not give the same value when averaged over $\Omega(N)$
and over $\Omega(N{-}1)$.
\subsection{Free volume}
\label{sec:fv}%
A well-established way for the above problem, passing from averages over
$N{-}1$~particles to $N$~particles, but without changing the system, is to
introduce the so-called \emph{free-volume}
averages~\cite{Speedy81,SpeRei91a,CorBow99,SasTruDebTorSti98}. This method
allows to measure pressures even in crystalline or nearly jammed systems.

From a configuration~$k\in\Omega(N)$ we choose one particle at~$\bfr_i$, take
it out, and such construct the reduced
state~$\takeout{k}{\bfr_i}\in\Omega(N{-}1)$. The resulting available
volume $V_0\bigl(\takeout{k}{\bfr_i}\bigr)$ is nonzero. One of its cavities
contains the point~$\bfr_i$. This cavity is called the \emph{free volume},
denoted by~$V_f\bigl(\takeout{k}{\bfr_i}\bigr)$. It is depicted in blue in
fig.~\ref{fig:av_fv}b. In terms of free volumes, eq.~\eqref{mono:p:av} reads
\begin{equation}
  \label{mono:p:fv}
  \frac{pV}{NkT} = 1 + \frac{\sigma}{2d}
    \biggl\langle\frac{S_f}{V_f}\biggr\rangle_{\calF(N)},
\end{equation}
where the ratio of averages has turned into the average of a ratio. The precise
definition of~$\calF(N)$ requires going into greater detail, which is done in
several steps in the remainder of this section. The passage from
\eqref{mono:p:av}~to \eqref{mono:p:fv} has been described several times in the
literature~\cite{Speedy81,SpeRei91a,CorBow99,SasTruDebTorSti98}. We find it
worth summarising these references in order to make the difference to our new
method clear, which will be introduced in sect.~\ref{sec:avato}.

In a first step, one constructs the set $\calC(N{-}1)$ of all possible cavities
which can be obtained from~$\Omega(N{-}1)$. The ratio of averages has not
changed at this point, and both are counting averages,
\begin{equation}
  \label{mono:p:cavityavgs}
  \frac{\bigl\langle S_0(k)\bigr\rangle_{k\in\Omega(N{-}1)}}
       {\bigl\langle V_0(k)\bigr\rangle_{k\in\Omega(N{-}1)}} =
  \frac{\bigl\langle S_c(k,l)\bigr\rangle_{(k,l)\in\calC(N{-}1)}}
       {\bigl\langle V_c(k,l)\bigr\rangle_{(k,l)\in\calC(N{-}1)}}.
\end{equation}
Here, we labelled every cavity by both the configuration~$k$ in which it occurs
and a number $l=1,\dots,N_c(k)$, where $N_c(k)$~is the number of cavities in
this configuration. $V_c(k,l)$ denotes the volume of the cavity, and $S_c(k,l)$
its boundary.

The next step introduces a \emph{probabilistic} element in the averages used in
eqs.~\ref{mono:p:cavityavgs}. Instead of averaging over all possible cavities,
one chooses cavities at random with a probability proportional to their
volume~$V_c(k,l)$~\cite{SasTruDebTorSti98},
\begin{equation}
  P(k,l) := V_c(k,l) \biggm/ \sum_\scriptbox{3em}{(k',l')\in\calC(N{-}1)} V_c(k',l').
\end{equation}
The corresponding average of a quantity~$f(k,l)$ is denoted by
\begin{equation}
  \bigl\langle f \bigr\rangle_{\calP(N{-}1)}
    := \sum_\scriptbox{3em}{(k,l)\in\calC(N{-}1)} P(k,l) f(k,l).
\end{equation}
The ratio of averages in eq.~\eqref{mono:p:av} now becomes an average of ratios,
\begin{equation}
  \frac{\bigl\langle S_c(k,l)\bigr\rangle_{\calC(N{-}1)}}
       {\bigl\langle V_c(k,l)\bigr\rangle_{\calC(N{-}1)}} =
  \biggl\langle\frac{S_c(k,l)}{V_c(k,l)}\biggr\rangle_{\calP(N{-}1)}
\end{equation}
In a last step one passes from $(N{-}1)$ to $N$~spheres, which defines the
average over~$\calF(N)$ as the following procedure: One chooses uniformly a
configuration~$k\in\Omega(N)$ and then again uniformly one of the spheres
$i\in\{1,\ldots N\}$. This latter choice can be repeated many times without
changing the probability space, such that in the end it is equal to
deterministically choosing every sphere once. In order to prove the equivalence
of the averages over $\calF(N)$ and over $\calP(N{-}1)$, one has to show that
the cavity probability~$P(k',l)$ ($k'\in\Omega(N{-}1)$) leads indeed to a
homogeneous distribution of configurations~$k\in\Omega(N)$, and vice
versa.

On the algorithmic level, what eq.~\eqref{mono:p:fv} means in the analysis of a
series of snapshots is the following:\\
\begin{minipage}{\columnwidth}
\begin{algo}
\item Loop over the snapshots~$k\in\calS(N)$:
  \begin{algo}
  \item Loop over all particles~$i$:
    \begin{algo}
    \item Take out particle~$i$.
    \item Increase all other diameters by $\sigma_i$.
    \item Find the space not covered by any sphere (cavities).
    \item Identify the cavity which contains the center~$\bfr_i$.
    \item $V_f\bigl(\takeout{k}{\bfr_i}\bigr) \leftarrow$ volume of this cavity.
    \item $S_f\bigl(\takeout{k}{\bfr_i}\bigr) \leftarrow$ boundary of this cavity.
    \item Accumulate $S_f/V_f$ over both loops.
    \end{algo}
  \item End loop over~$i$.
  \end{algo}
\item End loop over~$k$.
\end{algo}
\end{minipage}
or as a formula,
\begin{equation}
  \label{mono:p:FV}
  \frac{pV}{NkT} - 1 = \frac{\sigma}{2d}
    \biggl\langle
    \frac{1}{N}\sum_{i=1}^N
    \frac{S_f\bigl(\takeout{k}{\bfr_i}\bigr)}
         {V_f\bigl(\takeout{k}{\bfr_i}\bigr)}\biggr\rangle_{k\in\calS(N)}
\end{equation}
We write eq.~\eqref{mono:p:FV} as an identity, which is strictly valid only in
the limit of an infinite number of snapshots.
\subsection{Available volume after take-out}
\label{sec:avato}%
We will now develop a third, new method which turns out to be an alternative to
the available-volume and free-volume averages.

Let us start with a naive algorithmic question: In the above algorithm, why not
take the whole available volume of the reduced state instead of only one of its
cavities? In the configuration of fig.~\ref{fig:av_fv}b we would take the union
of the blue and the red cavities. We call this union volume the \emph{available
volume after take-out} (\avato), denoted
by~$V_0\bigl(\takeout{k}{\bfr_i}\bigr)$. An average over this volume inherits
the advantage from the \fv-average that even at high densities there is always
at least one cavity to calculate. With the \av-average it shares the advantage
that there are more than one cavity, possibly many, which contribute when the
system is not dense. This improves the stability of the averages. In this sense
the \avato-average combines the best of two worlds.

Using this \avato-average, the pressure is calculated by
\begin{equation}
  \label{mono:p:avato}
  \frac{pV}{NkT} = 1 + \frac{\sigma}{2d}
    \biggl\langle
    \frac{1}{N}\sum_{i=1}^N
    \frac{S_0\bigl(\takeout{k}{\bfr_i}\bigr)}
         {V_0\bigl(\takeout{k}{\bfr_i}\bigr)}\biggr\rangle_{k\in\Omega(N)}
\end{equation}

The answer to the above question is affirmative: Equation~\eqref{mono:p:avato} is
entirely equivalent to eq.~\eqref{mono:p:av}. Even better, this equivalence is
based only on the counting of states, no additional probability is required.
Let us start with all configurations $k'\in\Omega(N{-}1)$. If a state~$k'$ is
extensible, it can be the result of taking one particle out of a configuration
$k\in\Omega(N)$. In fact, there are exactly $V_0(k')/\omega$ different such
states~$k$ which lead to~$k'$. If a state~$k'$ is inextensible, we have
$V_0(k')=0$. In the reverse direction, starting with all configurations
$k\in\Omega(N)$, for each of them there are exactly $N$~ways to produce an
extensible state~$k'$. The such constructed mapping between states~$k$ and $k'$
is far from being one-to-one, but the whole $\Omega(N)$ is mapped to the
extensible states, a subset of $\Omega(N{-}1)$. This means that for any
function~$f: \Omega(N{-}1)\to\Reals$ we have the equality
\begin{equation}
  \label{mono:p:link}
  \sum_{k\in\Omega(N)} \sum_{i=1}^{N} f\bigl(\takeout{k}{\bfr_i}\bigr)
  = \sum_\scriptbox{3em}{k'\in\Omega(N{-}1)} \frac{V_0(k')}{\omega}\: f(k').
\end{equation}
Every extensible state is counted an equal number on both sides of the equation.
Notice that the weighting $V_0(k')/\omega$ automatically eliminates the
inextensible states from the sum on the right-hand side. We may therefore sum
over all states~$\Omega(N{-}1)$. Notice further that when we specialize
eq.~\eqref{mono:p:link} to the constant function $f(k')=1$, we obtain Speedy's
eq.~(5) for the number of configurations~\cite{Speedy88}. The equivalence of
eqs.~\eqref{mono:p:av} and~\eqref{mono:p:avato} is now shown by writing out the
averages as sums and then using eq.~\eqref{mono:p:link} twice, with
$f(k')=S_0(k')/V_0(k')$ and with $f(k')=1$.

When applied to a finite set of snapshots, only the type of average changes, as
compared to eq.~\eqref{mono:p:avato},
\begin{equation}
  \label{mono:p:AVATO}
  \frac{pV}{NkT} - 1 = \frac{\sigma}{2d}
    \biggl\langle
    \frac{1}{N}\sum_{i=1}^N
    \frac{S_0\bigl(\takeout{k}{\bfr_i}\bigr)}
         {V_0\bigl(\takeout{k}{\bfr_i}\bigr)}\biggr\rangle_{k\in\calS(N)}
\end{equation}
Again, this approximation becomes exact in the limit of infinite snapshots. The
corresponding algorithm is nearly identical to the one for free volumes, only
the averaged quantity is different,\\
\begin{minipage}{\columnwidth}
\begin{algo}
\item Loop over the snapshots~$k\in\calS(N)$:
  \begin{algo}
  \item Loop over all particles~$i$:
    \begin{algo}
    \item Take out particle~$i$.
    \item Increase all other diameters by $\sigma_i$.
    \item Find the space not covered by any sphere (cavities).
    \item $V_0\bigl(\takeout{k}{\bfr_i}\bigr) \leftarrow$ sum of all cavity volumes.
    \item $S_0\bigl(\takeout{k}{\bfr_i}\bigr) \leftarrow$ sum of all cavity boundaries.
    \item Accumulate $S_0/V_0$ over both loops.
    \end{algo}
  \item End loop over~$i$.
  \end{algo}
\item End loop over~$k$.
\end{algo}
\end{minipage}
%
\subsection{Polydisperse case: pressure}
\label{sec:press}

For simplicity of the notation, we derive the expressions for a binary mixture
only, the generalisation to multicomponent systems is evident from the
structure of the formulae. The mixture contains $N_A$~spheres of
diameter~$\sigma_A$ and $N_B$~spheres of a different diameter~$\sigma_B$. In a
configuration
\begin{equation}
  k = \Bigl( \bigl\{\cA_1,\dots\cA_{N_A}\bigr\}, \bigl\{\cB_1,\dots\cB_{N_B}\bigr\} \Bigr)
\end{equation}
all the $A$-particles can be interchanged among themselves without changing the
state, and equally the $B$-particles. The set of all possible states is denoted
by~$\Omega(N_A,N_B)$. When we allow particles to have different sizes, also the
available volume starts to depend on the (type of) particle. We will write
$V^\alpha_0$ for the volume where we can insert the center of another particle
\emph{of type~$\alpha$}, and accordingly for $S^\alpha_0, V^\alpha_f, \dots$

In order to generalize eq.~\eqref{mono:p:av} to the polydisperse case, we can
follow the same arguments as in Ref.~\cite{Speedy88}. This is a straightforward
task, but it has not yet been published. 
The result is similar to Corti~\& Bowles' eq.~(59), only that the $(N{-}1)$
averages show up explicitly,
\begin{equation}
  \label{poly:p:av}
  \frac{pV}{N\,kT} - 1 = \frac{1}{2dN} \sum_{\alpha=A,B} N_\alpha\sigma_\alpha
      \frac{\bigl\langle S^\alpha_0\bigr\rangle_{\Omega(N_\alpha-1,N_\cdot)}}%
           {\bigl\langle V^\alpha_0\bigr\rangle_{\Omega(N_\alpha-1,N_\cdot)}}.
\end{equation}
The notation $\Omega(N_\alpha{-}1,N_\cdot)$ stands for $\Omega(N_A{-}1,N_B)$ or
for $\Omega(N_A,N_B{-}1)$, respectively. The algorithm now includes a loop over
the particle types,\\
\begin{minipage}{\columnwidth}
\begin{algo}
\item Loop over the snapshots~$k\in\calS(N)$:
  \begin{algo}
  \item Loop over the particle types~$\alpha$:
    \begin{algo}
    \item Enlarge all sphere diameters by~$\sigma_\alpha$.
    \item Find the space not covered by any sphere (cavities).
    \item $V^\alpha_0(k) \leftarrow$ sum of all cavity volumes.
    \item $S^\alpha_0(k) \leftarrow$ sum of all cavity boundaries.
    \item Accumulate $S^\alpha_0(k)$ and $V^\alpha_0(k)$ in averages.
    \end{algo}
  \item End loop over~$\alpha$.
  \end{algo}
\item End loop over~$k$.
\end{algo}
\end{minipage}
The algorithm evaluates the average in the available-volume approximation for
the pressure:
\begin{equation}
  \label{poly:p:AV}
  \tag{$p$\discretionary{/}{/}{/}\av}
  \frac{pV}{N\,kT} - 1 \approx \frac{1}{2dN} \sum_{\alpha=A,B} N_\alpha\sigma_\alpha
      \frac{\bigl\langle S^\alpha_0(k)\bigr\rangle_{k\in\calS(N_A,N_B)}}%
           {\bigl\langle V^\alpha_0(k)\bigr\rangle_{k\in\calS(N_A,N_B)}}
\end{equation}

From the available-volume average in eq.~\eqref{poly:p:av} we can follow the
same arguments as in sect.~\ref{sec:fv} to obtain the free-volume average,
\begin{equation}
  \label{poly:p:fv}
  \frac{pV}{N\,kT} - 1 = \frac{1}{2dN} \sum_{\alpha=A,B} N_\alpha\sigma_\alpha
  \biggl\langle\frac{S^\alpha_f}{V^\alpha_f}\biggr\rangle_{\calF^\alpha(N_A,N_B)}
\end{equation}
which generalises eq.~\eqref{mono:p:fv} to the polydisperse case. This is
Corti~\& Bowles' eq.~(75). In practice, the free volumes are different
geometrical objects for each~$i$ and each~$\alpha$, such that it is simpler to
write one sum over all particles instead of separate sums over types and
particles. We had this sum already in eq.~\eqref{mono:p:FV}, which now becomes
\begin{equation}
  \label{poly:p:FV}
  \tag{$p$\discretionary{/}{/}{/}\fv}
  \frac{pV}{NkT} - 1 = \frac{1}{2dN}
    \sum_{i=1}^N \sigma_i
    \biggl\langle
    \frac{S^{\alpha_i}_f\bigl(\takeout{k}{\bfr_i}\bigr)}
         {V^{\alpha_i}_f\bigl(\takeout{k}{\bfr_i}\bigr)}\biggr\rangle_\scriptboxl{0pt}{k\in\calS(N_A,N_B)}
\end{equation}
Here, $\sigma_i$ and $\alpha_i$ are diameter and type of particle~$i$, respectively.

Finally, the arguments of sect.~\eqref{sec:avato} hold for every particle type
individually. For each~$\alpha$ we obtain an equation such
as~\eqref{mono:p:link},
\begin{equation}
  \label{poly:p:link}
  \sum_{k\in\Omega(N_A,N_B)} \sum_{i=1}^{N_\alpha} f^\alpha\bigl(\takeout{k}{\bfr^\alpha_i}\bigr)
  = \sum_\scriptbox{3em}{k'\in\Omega(N_\alpha{-}1,N_\cdot)} \frac{V^\alpha_0(k')}{\omega}\: f^\alpha(k'),
\end{equation}
where now also the averaged quantity depends on~$\alpha$. Using the
\avato~average, the pressure evaluation from data then turns out to be
\begin{equation}
  \label{poly:p:AVATO}
  \tag{$p$\discretionary{/}{/}{/}\avato}
  \frac{pV}{NkT} - 1 = \frac{1}{2dN}
    \sum_{i=1}^N \sigma_i
    \biggl\langle
    \frac{S^{\alpha_i}_0\bigl(\takeout{k}{\bfr_i}\bigr)}
         {V^{\alpha_i}_0\bigl(\takeout{k}{\bfr_i}\bigr)}\biggr\rangle_{\scriptboxl{0em}{k\in\calS(N_A,N_B)}}
\end{equation}
The generalisation of the analytical formula~\eqref{mono:p:avato} to the
polydisperse case is analogous, only that the average is done
over~$\Omega(N_A,N_B)$.

\subsection{Polydisperse case: chemical potentials}
\label{sec:mu}%
In continuum thermodynamics, the chemical potentials $\mu_A, \mu_B$ are said to
be derivatives of the suitable thermodynamic potentials with respect to $N_A,
N_B$. As these latter take discrete values, we have to choose either the upper
or the lower difference. In agreement with Ref.~\cite{SpeRei91a} we choose the
lower one, because it leads to averages over configurations of
$(N{-}1)$~particles, consistent with those in eq.~\eqref{mono:p:av}. In a
\emph{canonical} ensemble we have the lower difference of Helmholtz' free
energies,
\begin{equation}
  \label{mu:canon}
  \mu_A(T, V, N_A, N_B) :=
  \begin{aligned}[t]
    &{}+ F(T,N_A,N_B,V,\sigma_A,\sigma_B)\\
    &{}- F(T,N_A{-}1,N_B,V,\sigma_A,\sigma_B)
  \end{aligned}
\end{equation}
In the \emph{microcanonical} ensemble we have differences of entropies,
\begin{equation}
  \label{mu:micro}
  \frac{\mu_A}{T}(E, V, N_A, N_B) :=
  \begin{aligned}[t]
    &{}- S(E,N_A,N_B,V,\sigma_A,\sigma_B) \\
    &{}+ S(E,N_A{-}1,N_B,V,\sigma_A,\sigma_B).
    \end{aligned}
\end{equation}

The cavity methods analyse only the configuration integral and disregard the
kinetic part of the phase space. We thus require the phase space integral to
factorise into two parts -- which is the case both for the canonical partition
function (always), and in the considered hard-sphere case also for the
microcanonical integral. If we denote as $\varphi_\text{kin},
\varphi_\text{conf}$ the two factors of the phase space integral (synonymously
for the canonical and microcanonical cases), the differences of
eqs.~\eqref{mu:canon} and \eqref{mu:micro} become (written for particle type~$A$),
\begin{align}
  \notag
  \frac{\mu_A}{kT}
   &= \ln\biggl(\frac{\varphi_\text{kin}(N_A{-}1,N_B)}{\varphi_\text{kin}(N_A,N_B)}\:
                \frac{\varphi_\text{conf}(N_A{-}1,N_B)}{\varphi_\text{conf}(N_A,N_B)}
   \biggr) \\
  \notag
   &= \ln\biggl(\frac{\lambda_A^d}{\ell^d}\: \frac{|\Omega(N_A{-}1,N_B)|\ell^d}{|\Omega(N_A,N_B)|\omega}\biggr) \\
   \label{poly:mu:av}
   &= \ln\frac{\lambda_A^d\,N_A}{\bigl\langle V^A_0\bigr\rangle_{\Omega(N_A{-}1,N_B)}}.
\end{align}
Here, $\ell$ denotes an arbitrary length scale. The constants $\lambda_\alpha$
are the thermal de-Broglie wavelengths, given in the canonical case by
$h/\sqrt{2\pi\,kT\,m_\alpha}$, and in the microcanonical case by the indicated
ratio of $\varphi_\text{kin}$. The~$\lambda_\alpha$ only set the origin of the
energy axis and play no role in the following. Equation~\eqref{poly:mu:av} is
Corti~\&~Bowles' eq.~(48), only that here the average over $(N{-}1)$~particles
becomes explicit. In the last step we made use of their relation~(30).

The appearance of the $\Omega(N{-}1)$-average in eq.~\eqref{poly:mu:av} gives
rise to free-volume and \avato~averages, similar to the treatment of the
pressure in the above sections -- with the important difference, however, that
here we do not treat a relative property (a ratio such as $S/V$), but an
absolute one. This will hinder us from eliminating the average
$\langle\cdot\rangle_{\Omega(N{-}1)}$ entirely~\footnote{This \emph{caveat}
also leads to the problem mentioned by Sastry \emph{et
al.}~\cite{SasTruDebTorSti98}, that \emph{``the chemical potential cannot be
determined from free-volume information alone''}, we need also the number of
cavities.}. Concerning the average over~$\Omega(N{-}1)$, we have the same
problem as for the pressure, that we cannot calculate eq.~\eqref{poly:mu:av}
from data snapshots. The best we can do for the moment is to use the
statistical averages we can compute, arriving at
\begin{equation}
  \label{poly:mu:AV}
  \tag{$\mu$\discretionary{/}{/}{/}\av}
  \frac{\mu_\alpha}{kT}-\ln\lambda_\alpha^d
  \approx \ln\frac{N_\alpha}{\bigl\langle V^\alpha_0(k)\bigr\rangle_{k\in\calS(N_A,N_B)}}.
\end{equation}
This approximation aims at evaluating $\mu(N+1)$ instead of $\mu(N)$, as we saw
already for the pressure in eq.~\eqref{mono:p:AV}. Again, the approximation
breaks down for systems so dense that one cannot insert another particle.

The free-volume equivalent of eq.~\eqref{poly:mu:av} is
\begin{equation}
  \label{poly:mu:fv}
  \frac{\mu_A}{kT} - \ln\lambda_A^d
  = \ln\frac{N_A \bigl\langle 1/V^A_f\bigr\rangle_{\calF^A(N_A,N_B)}}%
            {\bigl\langle N^A_c(k)\bigr\rangle_{k\in\Omega(N_A{-}1,N_B)}}
\end{equation}
This is Sastry's eq.~(9)~\cite{SasTruDebTorSti98}, generalised to the
polydisperse case. Notice the average number of cavities in the denominator,
which still uses the counting average. When we want to turn this equation into
an algorithm, again, we have to replace $\bigl\langle
N^A_c(k)\bigr\rangle_{k\in\Omega(N_A{-}1,N_B)}$ by something we can compute,
for example $\bigl\langle N^A_c(k)\bigr\rangle_{k\in\calS(N_A,N_B)}$, without
knowing the error we make. Thus, in data analysis we will use
\begin{equation}
  \label{poly:mu:FV}
  \tag{$\mu$\discretionary{/}{/}{/}\fv}
  \frac{\mu_\alpha}{kT} - \ln\lambda_\alpha^d \approx
  \ln\frac{\sum\limits_{i=1}^{N_\alpha}\displaystyle \biggl\langle \frac{1}{V^\alpha_f\bigl(\takeout{k}{\bfr^\alpha_i}\bigr)} \biggr\rangle_{k\in\calS(N_A,N_B)}}%
          {\displaystyle \bigl\langle N_c^\alpha\bigr\rangle_{\calS(N_A,N_B)}}
\end{equation}
Upon increasing density, we will be limited by the denominator, just as in
eq.~\eqref{poly:mu:AV}. As soon as most of the states are not extensible
anymore, the average number of cavities vanishes.

Finally, let us calculate the chemical potentials in terms of \avato~averages. Here, the difference between extensible and inextensible states is
essential. We therefore introduce the notation $\Omega^{+}(N_A{-}1,N_B)$ for
those states in $\Omega(N_A{-}1,N_B)$ which are extensible by a particle of
type~$A$. Equation~\eqref{poly:p:link} does not work with quantities such as
$f(k')=1/V_0^\alpha(k')$ which are infinite for inextensible states. They would
formally annihilate the weighting $V_0^\alpha(k')$ which is necessary to
discard inextensible states from the sum. Using the restricted summation
repairs this problem,
\begin{multline}
  \sum_{k\in\Omega(N_A,N_B)} \sum_{i=1}^{N_A} \frac{1}{V_0^A\bigl(\takeout{k}{\bfr^A_i}\bigr)}
  = \sum_\scriptbox{5em}{k'\in\Omega^{+}(N_A{-}1,N_B)} 1/\omega \\
  = \frac{\bigl|\Omega^{+}(N_A{-}1,N_B)\bigr|}{\omega}.
\end{multline}
The take-out average of $f^A=1/V^A_0$ becomes
\begin{multline}
  \biggl\langle \frac{1}{V^A_0} \biggr\rangle_{\calT^A(N_A,N_B)}
 := \frac{1}{N_A}\sum_{i=1}^{N_A} \biggl\langle
    \frac{1}{V^A_0\bigl(\takeout{k}{\bfr^A_i}\bigr)}
    \biggr\rangle_{k\in\Omega(N_A,N_B)} \\
  = \frac{\bigl|\Omega^{+}(N_A{-}1,N_B)\bigr|}%
         {\sum\limits_\scriptbox{3em}{k'\in\Omega^{+}(N_A{-}1,N_B)} V^A_0(k')}
  = \frac{1}{\gamma_A}\:
    \frac{1}{\bigl\langle V^A_0\bigr\rangle_{\Omega(N_A{-}1,N_B)}}.
\end{multline}
In the last step we introduced the ratio of the total number of states and the
number of extensible states,
\begin{equation}
  \gamma_A :=  \frac{\bigl|\Omega(N_A{-}1,N_B)\bigr|}%
                    {\bigl|\Omega^{+}(N_A{-}1,N_B)\bigr|}.
\end{equation}
The chemical potentials are then written as
\begin{equation}
  \label{poly:mu:avato-a}
  \frac{\mu_\alpha}{kT} - \ln\lambda_\alpha^d
   = \ln N_\alpha \gamma_\alpha
     \biggl\langle \frac{1}{V^\alpha_0} \biggr\rangle_\scriptboxl{2em}{\calT^\alpha(N_A,N_B)}
\end{equation}
The occurrence of the ratio~$\gamma_\alpha$ in this equation is unfortunate. We
will not be able to compute it from $N$-particle snapshots. In dilute systems,
where we can insert another particle in practically all configurations, this
factor is close to unity and does not interfere. In dense systems, however, we
do not know in advance into how many configurations we can insert a particle,
knowing only that the factor will be larger than one, diverging for the largest
possible density. The following approximation therefore \emph{underestimates}
the true chemical potential:
\begin{equation}
  \label{poly:mu:AVATO-A}
  \tag{$\mu$\discretionary{/}{/}{/}\avatoa}
  \frac{\mu_\alpha}{kT} - \ln\lambda_\alpha^d
    \lessapprox
    \ln \sum_{i=1}^{N_\alpha} \biggl\langle
      \frac{1}{V^\alpha_0\bigl(\takeout{k}{\bfr^\alpha_i}\bigr)}
      \biggr\rangle_\scriptboxl{2em}{k\in\calS(N_A,N_B)}
\end{equation}

We may try to find different expressions, on the search to avoid the problem
with the~$\gamma_\alpha$ in eq.~\eqref{poly:mu:avato-a}. For example, we may
average $N^A_c/V^A_0$, which is a relative property, thus $\gamma_A$~will not
intervene. Its take-out average is
\begin{multline}
  \biggl\langle \frac{N^A_c}{V^A_0} \biggr\rangle_{\calT^A(N_A,N_B)}
  = \frac{\bigl\langle N^A_c\bigr\rangle_{\Omega(N_A{-}1,N_B)}}%
         {\bigl\langle V^A_0\bigr\rangle_{\Omega(N_A{-}1,N_B)}} \\
  = \biggl\langle \frac{1}{V^A_f} \biggr\rangle_{\calF^A(N_A,N_B)},
\end{multline}
we find a quantity which we have seen already, namely the free-volume average
of $1/V^A_f$. We can thus express eq.~\eqref{poly:mu:fv} in yet another way,
namely
\begin{equation}
  \label{poly:mu:avato-b}
  \frac{\mu_A}{kT} - \ln\lambda_A^d
   = \ln\frac{N_A \bigl\langle N^A_c/V^A_0\bigr\rangle_{\calT^A(N_A,N_B)}}%
             {\bigl\langle N^A_c\bigr\rangle_{\Omega(N_A{-}1,N_B)}}
\end{equation}
Of course, this expression suffers from the same limitations concerning the
denominator as does eq.~\eqref{poly:mu:fv}. It is nevertheless interesting to
see how we can again replace an average of free volumes by an \avato~average.
In data analysis, we will use the following approximation for the denominator
in eq.~\eqref{poly:mu:avato-b},
\begin{multline}
  \label{poly:mu:AVATO-B}
  \tag{$\mu$\discretionary{/}{/}{/}\avatob}
  \frac{\mu_\alpha}{kT} - \ln\lambda_\alpha^d
  \approx
  \ln N_\alpha\frac{\bigl\langle N^\alpha_c/V^\alpha_0\bigr\rangle_{\calT^A(N_A,N_B)}}%
                   {\bigl\langle N^\alpha_c\bigr\rangle_{\calS(N_A,N_B)}} \\
  = \ln\frac{\sum\limits_{i=1}^{N_\alpha}\biggl\langle\frac{\textstyle N^\alpha_c(\takeout{k}{\bfr^\alpha_i})}
                                                           {\textstyle V^\alpha_0(\takeout{k}{\bfr^\alpha_i})}\biggr\rangle_{k\in\calS(N_A,N_B)}}%
            {\bigl\langle N^\alpha_c\bigr\rangle_{\calS(N_A,N_B)}}.
\end{multline}
%

\section{Numerical method and results}
We now proceed with the comparison of equations \eqref{poly:p:AV},
\eqref{poly:p:FV}, and \eqref{poly:p:AVATO} for the pressure, and of the
equations \eqref{poly:mu:AV}, \eqref{poly:mu:FV}, \eqref{poly:mu:AVATO-A},
\eqref{poly:mu:AVATO-B} for the chemical potentials by applying them to data.
Some of these equations are approximations to the real thermodynamic quantity,
so we hope to get from their comparison some information about the quality of
these approximations. In order to avoid at best other sources of error, we
constrain ourselves to snapshots which come from simulations, where we are
limited only by the numerical precision in the centers and radii of the
particles.
\subsection{Numerical methods}
\label{sec:nummethods}%
To generate snapshots, we used event-driven molecular
dynamics~\cite{Rapaport04} with elastic collisions, such that the statistical
ensemble is the microcanonical one -- without the need for a thermostat. In
addition to energy conservation, also the total linear momentum is conserved
during the simulation. All our simulations use periodic boundary conditions to
avoid possible inaccuracies from cutting cavities at boundaries.

The snapshots contain $N=2150$~circular disks in two dimensions ($d=2$) which
have all identical mass $m_i=1$. The diameters of the disks are chosen randomly
from a Normal distribution with mean~$1$ and a prescribed standard deviation,
where outliers beyond three standard deviations were discarded. In different
simulations we used different standard deviations, starting with $0\%$
(monodisperse) and increasing in steps of $1.5\%$ until~$9\%$. Simulation units
are fixed by the average particle diameter, the particle mass and the
temperature ($kT=1$). At the beginning of a simulation, we shrunk the particles
drastically and arranged them on a hexagonal grid, with random velocities chosen
according to a Normal distribution -- subject to the two constraints of zero
total momentum and prescribed total energy. In a first phase of the simulation
run, the particles, being in a gas phase, were successively grown to their
individually prescribed target diameter. This was done without creating
additional collisions, such that the total energy and momentum remained
unaffected. After all diameters were attained, the system was allowed to relax
to its thermodynamic equilibrium in a second phase of the simulation run. The
equilibrium was either a gas state, at surface fractions $\phi < 0.70$, or a
crystalline state, at $\phi\gtrsim 0.72$, see footnote.\footnote{We neither discuss the
hexatic phase here, nor the precise location and nature of the phase
transitions.} Polydispersity shifted the transition to higher values: For
example, at $6\%$ we found the crystalline phase at $\phi\gtrsim 0.78$, with
disclination defects, and at $9\%$ polydispersity we did not find a crystalline
structure anymore. Finally, the simulation run was continued, and snapshots
were recorded periodically in time. The period was chosen such that on average
every particle collided five times between subsequent snapshots.

For the pressure, which is at the same time a thermodynamic and a mechanical
quantity, we have a reference value at our disposition. During simulation runs,
also the mechanical definition of pressure, that is the \emph{volume-averaged
linear momentum current} was recorded as a time average. It comprises a kinetic
part and a virial part~\cite{Forster90,Schindler10}. This pressure, which
converges very quickly because it is calculated from all collisions, will serve as
a reference in the comparison of the different cavity results. We verified that
this numerical reference pressure coincides with the known virial
expansion~\cite{sklogwiki,CliMcC06} and found deviations smaller than 0.1\% up
to $\phi=0.35$.\medskip
\begin{figure}
  \centering
  \includegraphics{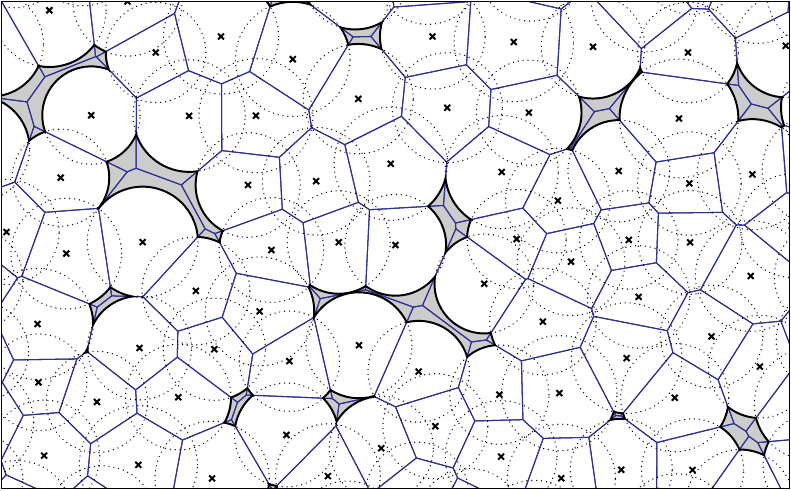}%
  \caption{The Voronoi cells (blue) which served to calculate the cavities in
  fig.~\ref{fig:av_fv}a. Dotted circles are the particles, extended by the
  radius of the particle for which the cavities are calculated.}%
  \label{fig:voronoi}
\end{figure}%

For measuring the volume and boundary of the cavity in all three methods, we
implemented the algorithm of Ref.~\cite{SasCorDebSti97}, see also
Refs.~\cite{SasTruDebTorSti98,MaiLakSas13}. We adapted it to the periodic case,
in two dimensions, for configurations where every particle can in principle
have a different diameter. The general idea of the algorithm is to cut space
into triangles, such that for each of them only a single disk is to be
considered. The triangles in question are pieces of the Voronoi cells which
come from a \emph{radical-plane construction}, see fig.~\ref{fig:voronoi} for
an example. The Voronoi diagram (also called power diagram) is the dual of a
\emph{weighted Delaunay} triangulation (also called \emph{regular
triangulation}), where the weights are the squares of the extended
radii~\cite{GelFin82,MT-cclrtw-redtp-10}. For the calculation of the regular
triangulation, we relied on the robust and efficient C++~library
CGAL~\cite{CGAL,MT-ct-c3pt-09}. Their algorithm scales as $N\log N$ in the number of
particles (instead of $N^2$ as in Ref.~\cite{MaiLakSas13}).

The periodic version of the regular triangulation in CGAL is about to be
developed, so that we had to implement this part ourselves, combining what CGAL
offers for periodic and for regular nonperiodic
triangulations~\cite{cgal:y-t2-15a,cgal:k-pt2-13-15a}. For the unweighted Delaunay
triangulation, it has been proved~\cite{DolHus97} that the periodic
triangulation can be extracted from $3\times 3$ periodic copies of the initial
input. We found that this result also holds for weighted Delaunay
triangulations.

The source code of the program to calculate the cavities will be made publicly
available~\cite{homepage}.
\subsection{Precision of the methods, pressure}
\label{sec:num:p}%
\begin{figure}
  \centering
  \includegraphics{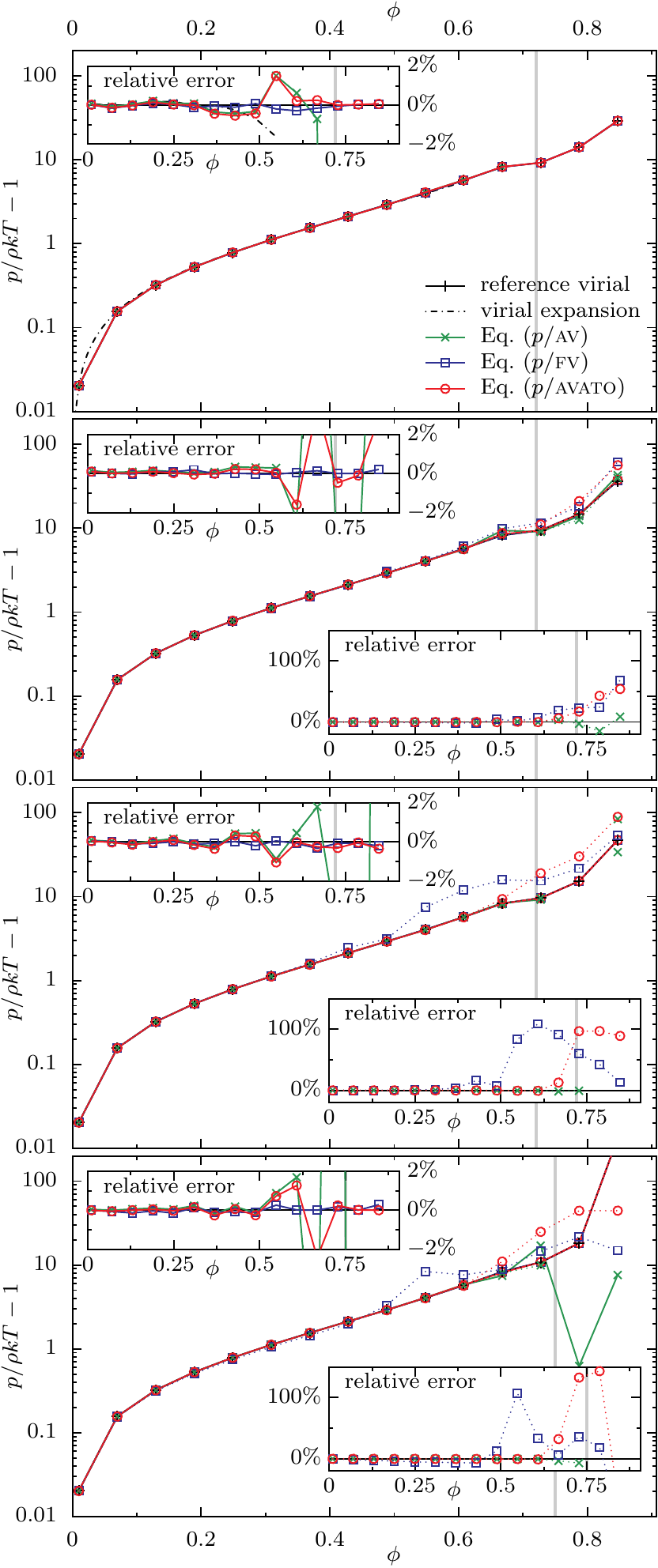}%
  \caption{Pressure extracted from molecular-dynamics snapshots. From top to
  bottom, polydispersities are $0\%$, $1.5\%$, $3\%$, and $6\%$. \textbf{Solid
  lines:}~the true radii are used; number of snapshots is $K=200$.
  \textbf{Dotted lines:}~disks are intentionally treated as if they were
  monodisperse; $K=1700$. Vertical gray lines are guides to the eye to separate
  liquid from crystalline states (without formal definition).}%
  \label{fig:poly:p}
\end{figure}%
The result of the data analysis for the pressure, according to
eqs.~\eqref{poly:p:AV}, \eqref{poly:p:FV}, and \eqref{poly:p:AVATO} are plotted
as solid lines in fig.~\ref{fig:poly:p}. We used 200~snapshots. Generally, all
three cavity methods coincide with the reference pressure within the linewidth,
only the \av~average shows extreme errors in the crystalline phase. This
behaviour is expected because snapshots which allow insertion of an ($N{+}1$)st
particle are very rare in the dense phase, resulting in an average over a
single cavity in the worst case.

In order to quantify the differences, the top-left insets of
fig.~\ref{fig:poly:p} show relative errors with respect to the reference
pressure. Generally speaking, the error is less than 1\%, with largest errors
in the dense gaseous phase, close to the transition to more ordered phases.
There, the \avato~average shows slightly larger fluctuations than the
\fv~average. Far in the crystalline phase the errors of the \fv{} and the
\avato~averages then drop again to very small values. The errors of the
\av~average grow strongly at around $\phi\gtrsim0.65$, again because the number
of extensible snapshots decreases strongly.
\subsection{Chemical potentials}
\label{sec:num:mu}%
\begin{figure}
  \centering
  \includegraphics{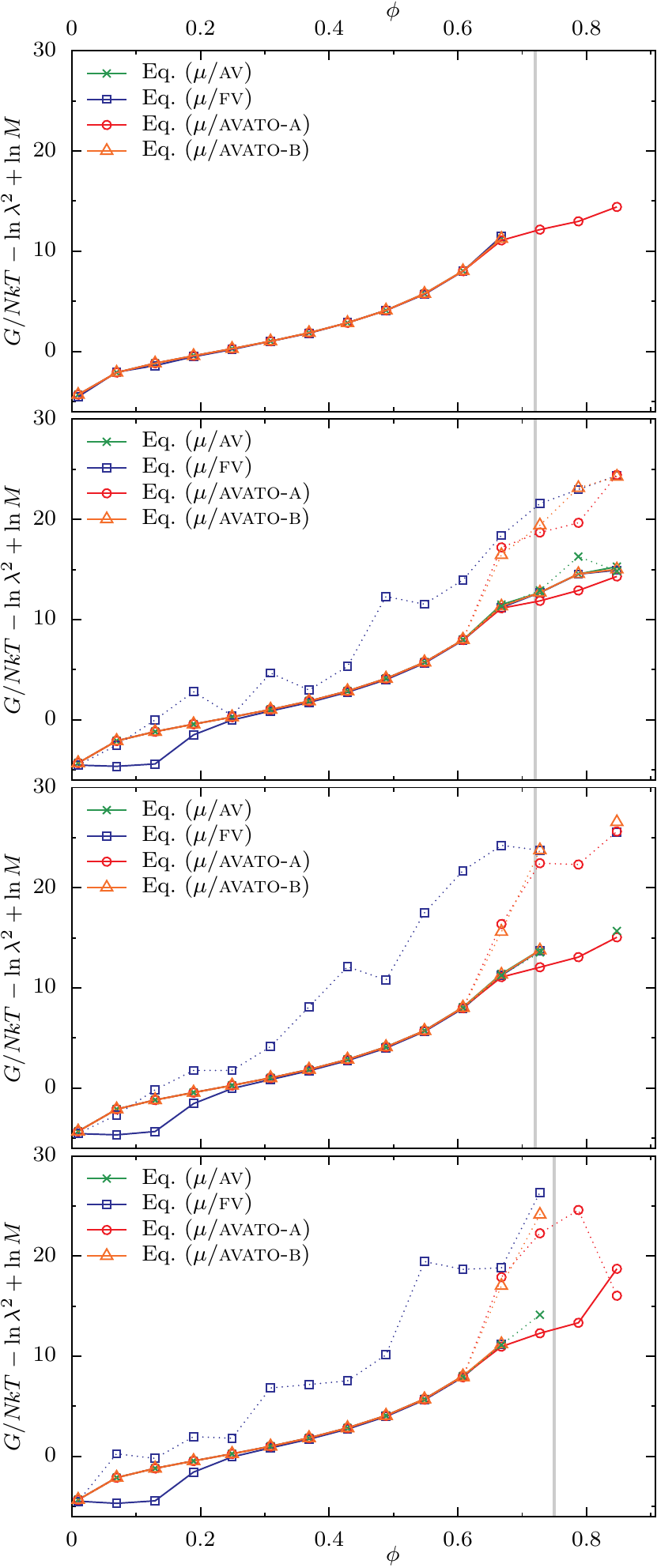}%
  \caption{Free enthalpy per particle, extracted from molecular dynamics
  snapshots. From top to bottom, polydispersities are $0\%$, $1.5\%$, $3\%$,
  and $6\%$. \textbf{Solid lines:}~the true radii are used; number of snapshots
  is $K=200$. \textbf{Dotted lines:}~disks are intentionally treated as if they
  were monodisperse; $K=1700$.}%
  \label{fig:poly:mu}
\end{figure}%

For the chemical potentials, which are not mechanical properties, we do not
have reference data at our disposition, and we can only plot the cavity
averages. Since a truly polydisperse system has as many particle classes as
particles, instead of all the chemical potentials we plot the free enthalpy,
also known as Gibbs free energy,
\begin{equation}
  \frac{G}{kT} = \sum_{\alpha=1}^M N_\alpha \frac{\mu_\alpha}{kT}.
\end{equation}
Here, $M$~denotes the number of classes; for $0\%$~polydispersity we have
$M=1$, otherwise $M=N$. In order to make free enthalpies with different numbers
of classes comparable, we had to add a term $\ln M$. The result of the data
analysis according to eqs.~\eqref{poly:mu:AV}, \eqref{poly:mu:FV},
\eqref{poly:mu:AVATO-A}, and \eqref{poly:mu:AVATO-B} are plotted as solid lines
in fig.~\ref{fig:poly:mu}. We used the same 200~snapshots as for the pressure.

In the figure, we see good agreement between the four averages, with an
exception of the \fv~average which shows too low energies at moderately small
surface fractions, $0.07\lesssim\phi\lesssim0.2$. We do not have a good
explanation for this deviation at the moment, but we observe that it is related
to the number of classes, thus stems from the denominator in
eq.~\eqref{poly:mu:FV}. The deviation disappears when using $M=1$ in the
monodisperse case in the top panel of fig.~\ref{fig:poly:mu}. It reappeared in
a test where we forced $M=N$ on the same data. However, the same denominator is
found also in the \avato~average in eq.~\eqref{poly:mu:AVATO-B}, where it does
not cause a deviation. We tested also a bi-disperse system and found pronounced
fluctuations in the \fv~average -- to smaller and larger values -- just at
those values of $\phi$ where it exhibits the too low energies in
fig.~\ref{fig:poly:mu}. We thus conclude that eq.~\eqref{poly:mu:FV} is
extremely sensible in this $\phi$-region and converges more slowly than
elsewhere.%
\begin{figure}
  \centering
  \includegraphics{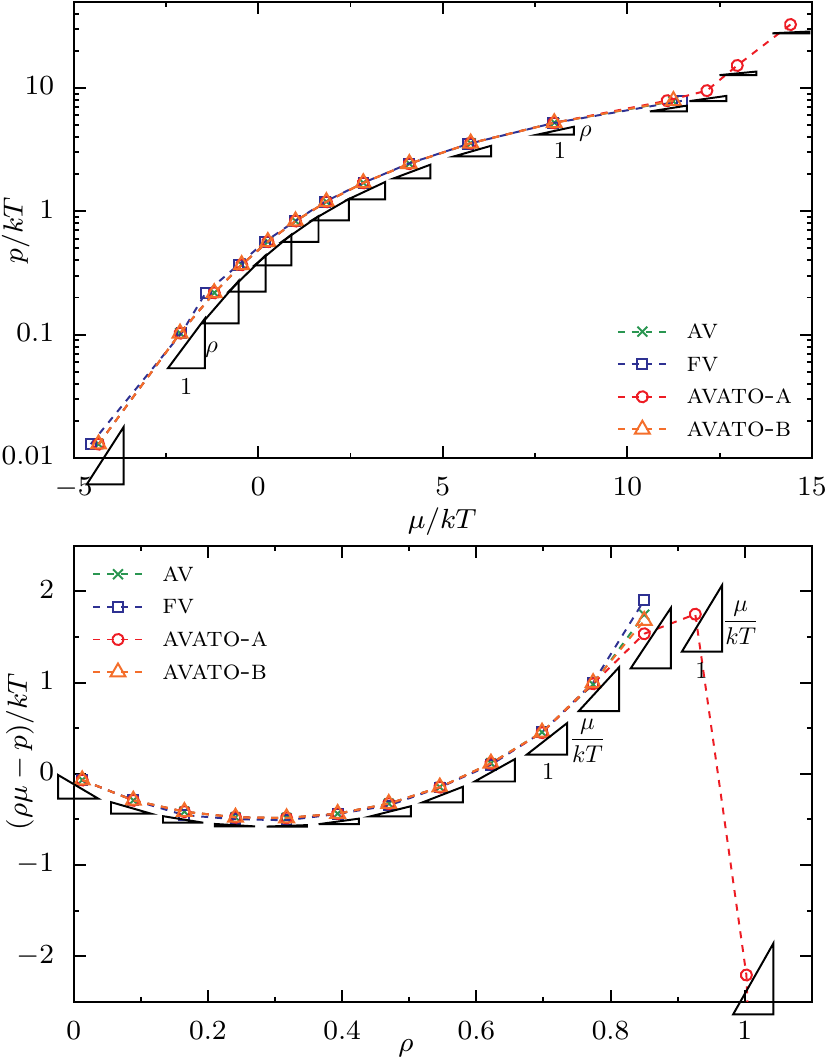}%
  \caption{Test of the Gibbs--Duhem relation on monodisperse data. We used the
  shortcut~$\mu/kT$ instead of $G/NkT-\ln\lambda^2+\ln M$. The little triangles
  indicate the expected local derivatives according to eqs.~\eqref{gd}.}%
  \label{fig:gd}
\end{figure}%

Furthermore, only one of the four methods, namely eq.~\eqref{poly:mu:AVATO-A}
gives values beyond the gaseous phase. The limitation for the three others
comes of course again from their denominators, which are \av~averages and thus
give no data if the snapshots are inextensible. The fourth method,
eq.~\eqref{poly:mu:AVATO-A} is known to underestimate systematically the
chemical potentials, so that one cannot rely on its results. We can check the
consistency of pressure and chemical potentials using the Gibbs--Duhem relation
$\sum_\alpha N_\alpha d\mu_\alpha = Vdp -SdT$, where the last term vanishes
because we worked at constant temperature. For the monodisperse case we rewrite
the relation in two ways which avoid numerical differentiation,
\begin{equation}
  \label{gd}
  \frac{dp}{d\mu} = \rho, \quad\text{and}\quad
  \frac{d(\rho\mu-p)}{d\rho} = \mu,
\end{equation}
where $\rho=N/V$ denotes the number density. Both variants of the equation are
plotted in fig.~\ref{fig:gd}. The little triangles indicate the expected
derivative, that is the right-hand side of the above equations. By eye, these
derivatives seem to coincide well with the general slope of the curves; only in
the crystalline part slope and triangles do not agree at all. This was expected
and proves that the values of eq.~\eqref{poly:mu:AVATO-A} beyond the gaseous
phase are not physical.
\subsection{Missing radius information}
\label{sec:num:missing}

Above, we have shown that the three averages, \av, \fv, and
\avato{} give similar answers when applied to numerical data. This is in
fact not surprising, since their equivalence has been established analytically.
The agreement of the above data is thus a test for the implementation and for
the convergence.

We go now a step further and investigate the robustness of the three averages
against imprecise snapshots. We will here consider only one source of
imprecision, that is lack of the individual radii of the disks/spheres. This is
a typical experimental situation, where one has a global idea of the radius
standard deviation, but where the resolution is insufficient to reliably
determine the radius of each particle~\cite{SchMarMegBry07}. The smallest
experimental radius standard deviation is about $4\%$~\cite{ZarNieSchBon13}, so
our question here is whether already this small polydispersity changes the
reliability of results extracted with one or the other cavity method.

On our numerical data this task can be done quite easily. It suffices to throw
away the information of individual radii. The algorithm which determines the
cavities works as well on freely invented radii as on the true ones. One could
for example attribute completely new radii according to a given distribution,
or shuffle the old ones among all particles. Generally, we think that the
details of radius attribution does not play a major role. We therefore follow a
conceptually simpler route, that is we treat the polydisperse system as if it
were monodisperse and attribute a unity diameter to every particle. The results
for the pressure and for the free enthalpy are plotted as dotted lines in
figs.~\ref{fig:poly:p} and~\ref{fig:poly:mu}.

For the pressure, we find extremely large deviations from the reference
pressure, which grow with the surface fraction~$\phi$. The \fv~average is the
first to deviate, it shows errors of~$100\%$ already far in the gas phase for
$\phi\gtrsim0.5$. One has to keep the polydispersity as tiny as~$1.5\%$ to keep
the errors of the \fv~average below~$20\%$. The \avato~average is more stable,
it works up to $\phi\approx0.7$. Remarkably both the \fv{} and the
\avato~averages are not robust against missing radius information in the
crystalline phase, not even for $1.5\%$~polydispersity -- Remember that their
errors were negligible when the radii were all correctly taken into account.
The third method, the \av~average works best up to the point where it fails
also for the full radii information. Generally speaking, also the fluctuations
in the curves is larger than in the full-radii case, although we took many more
snapshots into account.

For the chemical potentials, fig.~\ref{fig:poly:mu} shows a complete failure of
the \fv~method, even at rather small densities, and even at tiny
polydispersity. Errors are generally about 5--10\,kT, and the curves
fluctuate strongly. The \avato~average gives the correct result, until it deviates
also at high densities, $\phi\gtrsim0.66$. The \av~average, again, gives the
correct result in the $\phi$~range where it works.

\begin{figure}
  \centering
  \includegraphics{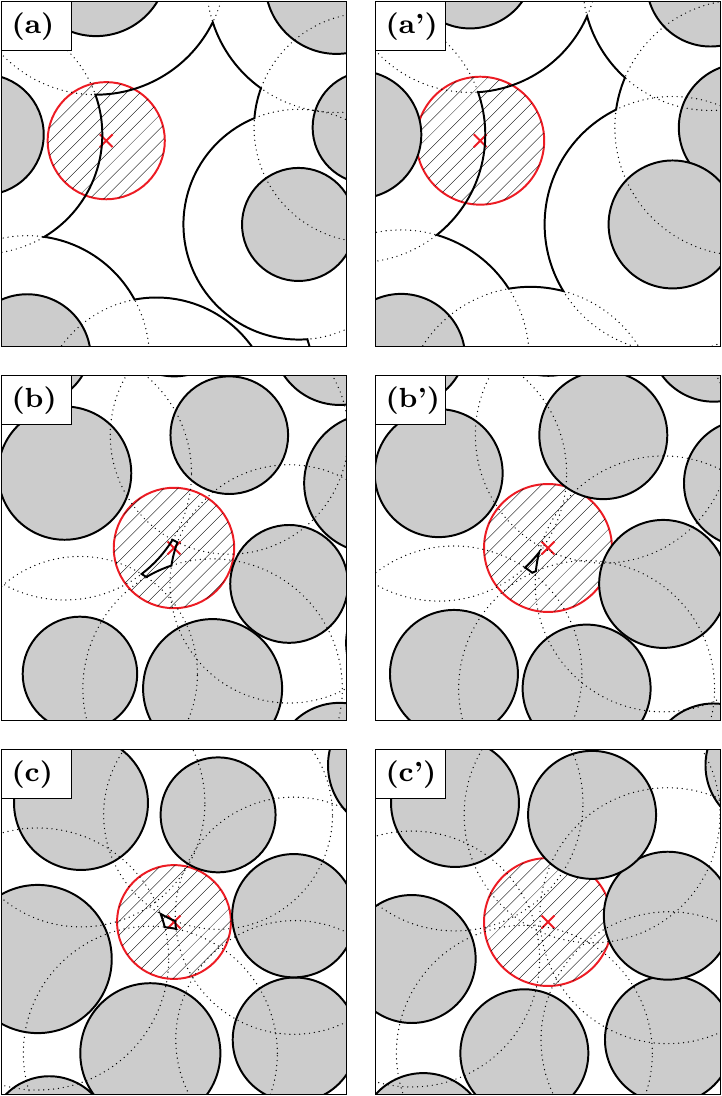}%
  \caption{Examples of what happens to the free-volume cavity in case of
  missing radius information. \textbf{Left}~with the true radii,
  \textbf{right}~with artificially identical radii. \textbf{(a', b')}~The
  particle center is not in ``the'' free-volume cavity. \textbf{(c')}~The
  cavity has vanished completely.}%
  \label{fig:loss}
\end{figure}%
In the search for an explanation why the \fv~method, and to some extend also
the \avato~method, are not robust against missing particle radii, we remark
that many configuration snapshots are \emph{incompatible} with a modified
radius attribution: Particles may overlap, as can be seen in the right-hand
column of fig.~\ref{fig:loss}. This has severe implications for the cavities,
especially for the free volumes. Figures~\ref{fig:loss}a',b' show cases where a
cavity, originally the one which was the free volume of the particle, is now
missed by the particle center. This cavity is not counted in the \fv~average,
but it is counted in the \avato~average. The smaller a cavity, the more easily
is it missed. However, especially the small cavities contribute strongly to the
chemical potentials, where $1/V$ is averaged. In the pressure, which averages
$S/V$, small cavities also contribute more strongly than large ones, but the scaling
is weaker. For very small cavities, a second effect occurs, they may completely
disappear, as shown in last row of fig.~\ref{fig:loss}. In total, it is
qualitatively understandable why we see more errors in dense than in dilute
systems, and why the errors are larger in chemical potentials than in
pressures, and why \fv~averages are more affected than the others.

We tested a number of variants of the original algorithms. Concerning missed
free volumes, we relaxed the criterion a bit and counted as ``free volume''
also cavities with a distance up to~$0.05$ from the particle center. The
cavities in figs.~\ref{fig:loss}a',b' would have been counted as ``free
volumes''. As another variant, we modified the normalisation of the averages.
In the original algorithm, it reads $\langle f \rangle_{\calS(N)} =
\frac{1}{K}\sum_{k\in\calS(N)} f(k)$. Instead of the total number of snapshots,
$K$, we now divided by the number of snapshots which gave a non-vanishing~$f$,
thus discarding those where cavities may have been missed. Finally, we modified
the number of particle ``classes'' ($M$) for the chemical potentials, grouping
the particles with similar radii together. None of the above variants, nor
combinations of them improved the dotted curves in figs.~\ref{fig:poly:p}
and~\ref{fig:poly:mu}.%

\section{Conclusions}
We investigated three geometrical methods to extract the pressure and the
chemical potentials from hard-sphere positions. All three methods are
cavity-based averages and differ in the precise way how the average is done and
what quantity is averaged. The theoretical derivation of these methods all
start with an average over all possibilities to place $(N{-}1)$~particles in a
given volume. The \fv~and \avato~methods consistently turn this average into
one over all possibilities to place $N$~particles in the volume, whereas the
\av~average stays with $(N{-}1)$~particles. The differences in how the average
is done and what quantity is averaged implies that the three methods differ in
their applicability to real-world data sets, that is in their convergence rate,
and in their response to noise.

As could be expected, there are hardly any problems in \emph{dilute, gas-like}
systems. Such systems easily explore their phase space, and it is easy to
insert another particle. Consequently, there is negligible difference in
averages over $N$~and over $(N{-}1)$~particles, and these averages tend to
converge well. We encountered, however, an exception to this general trend,
namely the combination \fv~average/chemical potentials in
fig.~\ref{fig:poly:mu}. With the same exception, we found that the methods are
robust to variations in the particle radii in the dilute regime, say for
$\phi\lesssim0.5$. This conclusion is in agreement with a previous
experimental/numerical study~\cite{DulKegAar08} done in three dimensions. In
this reference, the \av~average was calculated by counting pixels of the data
snapshots, a procedure which introduced errors of at least 5--10\% in the
individual particle radii, the system being minimally polydisperse. Further,
the precise position of the particles was limited by the Brownian diffusion
time scale. It appears that in the dilute regime, the \av~average -- together
with the additional data treatment described in the reference -- was
nevertheless able to give correct results for pressure and chemical potential.

Problems really start when systems become dense. At \emph{very high densities},
say $\phi\gtrsim0.75$, exploring the phase space takes a forbiddingly long
time, which, together with the uncertainty whether an average over $N$~and over
$(N{-}1)$ particles are equivalent, renders the \av~average unusable. Here, the
\fv~and \avato~methods propose a valuable alternative -- for calculating the
pressure. Indeed, we found excellent agreement of these two methods with the
reference pressure in fig.~\ref{fig:poly:p}. The chemical potentials, however,
are still out of reach, for the same reasons as for the \av~average.

Between dilute and very dense, there is a \emph{dense, but not extremely dense}
regime, say $0.5\lesssim\phi\lesssim0.75$. Depending on polydispersity and
initial preparation, such systems can be either (rather dilute) crystalline, or
disordered glass-like, or disordered with a very long relaxation time for
global order. These systems are in the main focus of many experimental studies
on colloids. In particular, one would like to really compare the ``free
energies'' of crystalline and disordered systems, and one thus requires both
the pressure and the chemical potentials. The question is of course, \emph{how
far can we push the cavity-based methods? Are they suitable to provide these
informations?} Despite the fact that this question is not directly the subject
of the present paper, we think that we provide some elements to its possible
answer. First, a close look on figs.~\ref{fig:poly:p} and~\ref{fig:poly:mu}
reveals that the \av~method can indeed give results close to the crystalline
phase, or even within. Its applicability thus has not a strict boundary but
rather gradually decreases together with the uncertainty whether an average
over $N$~and over $(N{-}1)$ particles are equivalent. With the \avato~average,
we here provide a second method which is at least conceptually consistent.
Comparison of \fv~and \avato{} results therefore reveals possible problems with
noise. The most direct information of our study on the above question is the
importance of high precision: An experimental method shall provide individual
radii with an error below 1\%~in order to allow proper extraction of
thermodynamic information on dense systems (see fig.~\ref{fig:poly:p}) -- a
challenging, but maybe not unreachable task.

Surely, every experiment has different sources of noise which are more or less
pertinent. We here provide an example for only one source of noise, or rather
of missing information, which was motivated by an existing experimental
situation~\cite{Daniel}. Experimentalists who are interested in calibrating
their data analysis are invited to use our cavity algorithm, which will be made
available as open source~\cite{homepage}. For those who do not want to
undertake heavy calibration, we propose the following rules of thumb: (i)~Avoid
the \fv~average. It is generally less robust against wrong radius information
than the other two methods. It has problems with chemical potentials at
moderately low densities. (ii)~If the density is below~$\phi\lesssim0.5$,
\avato~averages and \av~averages give the same result. \avato~averages are
conceptually cleaner, but depending on the number of classes, \av~averages
might be less expensive to calculate. (iii)~If your system is dense disordered
or crystalline, use more than one method and compare. (iv)~If you want to
implement only one method, stay with~\avato.

As a final remark, we would like to say that the algorithm is not restricted to
thermal systems. The physical interpretation, however, starts from the
assumption that all configurations are realised with the same probability. It
could therefore be interesting to apply the algorithms also to systems which
are not at equilibrium and where equivalence of all ensembles is not
guaranteed~\cite{CheEllZhaCheYunHenBriDauSaaLiuYod10}.

\begin{acknowledgement}
We thank Daniel Bonn and Rojman Zargar for interesting us for
the free-volume methods.
\end{acknowledgement}

\end{document}